\documentstyle[epsfig]{ioplppt}                % use this for journal style
\begin{document}
\jl{4}
\title{Test of high--energy interaction models using the hadronic core of EAS}[Test of high energy interaction models]

\author{

T.~Antoni, W.D.~Apel, K.~Bekk, K. Bernl\"ohr, E.~Bollmann, K.~Daumiller, 
P.~Doll, J.~Engler, F.~Fe{\ss}ler, H.J.~Gils, R.~Glasstetter, R.~Haeusler,
W.~Hafemann, A.~Haungs, D.~Heck, J.R.~H\"orandel 
\footnote{Corresponding author. FAX: +49--7247--82--4047, 
E--mail: joerg@ik1.fzk.de\\now at: University of Chicago, Enrico Fermi
Institute, Chicago, IL 60637}, T.~Holst, K.--H.~Kampert, H.O.~Klages, 
J.~Knapp
\footnote{now at: University of Leeds, Leeds LS2 9JT, U.K.},
H.J.~Mathes, H.J.~Mayer, J.~Milke, D.~M\"uhlenberg, J.~Oehlschl\"ager, 
H.~Rebel, M.~Risse, M.~Roth, G.~Schatz, H.~Schieler, F.K.~Schmidt, 
T.~Thouw, H.~Ulrich, J.~Unger
\footnote{now at: Hewlett--Packard GmbH, Herrenbergerstrasse,
D--71004 B\"oblingen}, J.H.~Weber, J.~Wentz, T.~Wiegert, D.~Wochele, 
J.~Wochele}
\address{
Institut f\"ur Kernphysik and Institut f\"ur Experimentelle 
Kernphysik, Forschungszentrum and Universit\"at Karlsruhe, 
P.O. Box 3640, D--76021 Karlsruhe, Germany}
\author{
J.~Kempa, T.~Wibig, J.~Zabierowski}
\address{
Institute for Nuclear Studies and Dept. of Experimental Physics, 
University of Lodz, PL--90950 Lodz, Poland}
\author{   
F.~Badea, H.~Bozdog, I.M.~Brancus, M.~Petcu, B.~Vulpescu}
\address{Institute of Physics and Nuclear Engineering, RO--7690 Bucharest, 
         Romania}
\author{ A.~Chilingarian, A. Vardanyan}
\address{Cosmic Ray Division, Yerevan Physics Institute, Yerevan 36, Armenia}

%
%  Uncomment out if preprint format required
%
%\pacs{00.00, 20.00, 42.10}
%\maketitle

\begin{abstract}
Using the large hadron calorimeter of the KASCADE experiment, 
hadronic cores of extensive air showers have been studied. The 
hadron lateral and energy distributions have been 
investigated in order to study the reliability of the shower 
simulation program CORSIKA with respect to particle transport, 
decays, treatment of low--energy particles, etc. 
A good description of the data
has been found at large distances from the shower core for several 
interaction models.
The inner part of the hadron distribution, on the other hand, 
reveals pronounced differences among interaction models. 
Several hadronic observables are compared with CORSIKA simulations
using the QGSJET, VENUS and SIBYLL models. QGSJET reproduces the hadronic
distributions best. At the highest energy, in the 10 PeV region, however, none 
of these models can describe the experimental data satisfactorily.
The expected number of hadrons in a shower is too large compared to the
observed number, when the data are classified according to the muonic
shower size.
\end{abstract}

%{\bf To be published in Journal of Physics G}

%\maketitle

\section{Proem}
The interpretation of extensive air shower (EAS) measurements in 
the PeV domain and above relies strongly on the hadronic interaction 
model applied when simulating the shower development in the 
Earth's atmosphere. Such models are needed to describe the 
interaction processes of the primary particles with the air nuclei 
and the production of secondary particles. 

In the EAS Monte Carlo codes the electromagnetic and weak 
interactions can be calculated with good accuracy. Hadronic 
interactions, on the other hand, are still uncertain to a large 
extent. A wealth of data exists on particle production from 
$p\overline{p}$ colliders up to energies which correspond to 
2~PeV/c laboratory momentum and from heavy ion experiments up to 
energies of 200~GeV/nucleon. However, almost all collider experiments 
do not register particles emitted in the very forward direction
where most of the energy flows. These particles carry the 
preponderant part of the energy and, therefore, are of utmost 
importance for the shower development of an EAS. Since most of 
these particles are produced in interactions with small momentum 
transfer, QCD is at present not capable of calculating their 
kinematic parameters.

Many phenomenological models have been developed to reproduce the 
experimental results. Extrapolations to higher energies, to small 
angles, and to nucleus--nucleus collisions have been performed under 
different theoretical assumptions. The number of participant 
nucleons in the latter case is another important parameter 
which influences the longitudinal development of a shower. 
Many EAS experiments have used specific models to determine the 
primary energy and to extract information about the primary mass 
composition. Experience shows that different models can lead to 
different results when applied to the same data. 

Therefore, it is of crucial importance to verify
the individual models experimentally as thoroughly as possible. 
When planning the KASCADE experiment, one of the principal 
motivations to build the hadron calorimeter was the intention to 
verify available interaction models by studying the hadronic 
central core. In the Monte Carlo code CORSIKA \cite{corsika} five different 
interaction codes have been implemented and placed at the users' 
disposal. By examining the hadron distribution in the very centre 
these interaction models are tested. The propagation code itself, viz. 
hadron transport, decay modes, scattering etc., is checked by 
looking to the hadron lateral distribution further outside up to 
distances of 100 m from the core. 

\section{The apparatus}
\begin{figure}\centering
 \epsfig{file=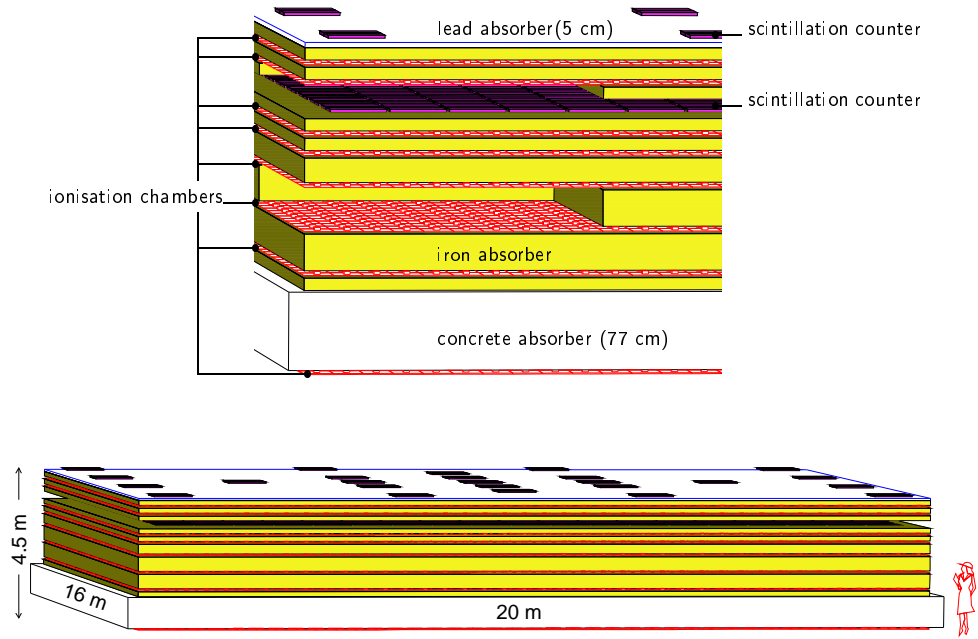,width=\textwidth,
	 clip=,bbllx=73,bblly=333,bburx=504,bbury=620}
 \caption{Sketch of the KASCADE central calorimeter. Detailed view (top)
 and total view (bottom).}
 \label{kalor}
\end{figure}
The KASCADE experiment consists of an array of 252 stations for 
electron and muon detection and a calorimeter in its centre for 
hadron detection and spectroscopy. 
It has been described in detail elsewhere \cite{kascade}. 
The muon detectors in the array are positioned directly below the 
scintillators of the electron detectors and are shielded by 
slabs of lead and iron corresponding to 20 radiation lengths in 
total. The absorber imposes an energy threshold of about 300 MeV for muon 
detection.

The calorimeter is of the sampling type, the energy being 
absorbed in an iron stack and sampled in eight layers by ionisation
chambers. Its performance is described in detail by Engler 
et al. \cite{kalorimeter}. 
A sketch of the set--up is shown in Fig.~\ref{kalor}. The iron 
slabs are 12--36~cm thick, becoming thicker in deeper parts of 
the calorimeter. Therefore, the energy resolution does not scale 
as $1/\sqrt{E}$, but is rather constant varying slowly from 
$\sigma/E = 20\%$ at 100~GeV to 10\% at 10~TeV. The concrete 
ceiling of the detector building is the last part of the 
absorber and the ionisation chamber layer below acts as tail catcher. 
In total, the calorimeter 
thickness corresponds to 11 interaction lengths $\lambda_{I}$ for 
vertical hadrons. On top, a 5~cm lead layer filters off the
electromagnetic 
component to a sufficiently low level.

The liquid ionisation chambers use the room temperature liquids 
tetramethylsilane (TMS) and tetramethylpentane (TMP). A detailed description 
of their performance can be found elsewhere \cite{prototyp}. Liquid ionisation 
chambers exhibit a linear signal behaviour with a very large dynamic range.
The latter is limited only by the electronics to about $5\times10^{4}$ 
of the amplifier rms--noise, i.e., the signal 
of one to more than $10^{4}$ passing muons, equivalent to 10 GeV 
deposited energy, is read out without saturation. This ensures 
the energy of individual hadrons to be measured linearly up to 
20\,TeV. At this energy, containment losses are at  a level of two 
percent. They rise and at 50 TeV signal losses of about 5\% 
have to be taken into account. The energy calibration is performed by 
means of through--going muons taking their energy deposition as standard. 
Electronic calibration is repeated in regular intervals of 6 months by
injecting a calibration charge at the amplifier input. A stability of 
better than 2\% over two years of operation has been attained. The 
detector signal is shaped to a slow signal with $10~\mu$s 
risetime in order to reduce the amplifier noise to a level less than
that of a passing muon. On the other hand, this makes a fast external 
trigger necessary. 

The principal trigger of KASCADE is formed by a coincident signal in at 
least five stations in one subgroup of 16 stations of the array.
This sets the energy threshold to a few times $10^{14}$~eV depending 
on zenith angle and primary mass. 
An alternative trigger is generated by a layer of plastic scintillators
positioned below the third iron layer at a depth of $2.2~\lambda_{I}$.
These scintillators cover two thirds of the calorimeter surface and
deliver
timing information with 1.5~ns  resolution.

\section{Simulations}
EAS simulations are performed using the CORSIKA versions 5.2 and 5.62 
as described in \cite{corsika}. The interaction models chosen in the 
tests are VENUS version 4.12 \cite{venus}, QGSJET \cite{qgsjet} and SIBYLL 
version 1.6 \cite{sibyll}. We have chosen two models which are based on the 
Gribov Regge theory because their solid theoretical ground allows best to 
extrapolate from collider measurements to higher  energies, forward 
kinematical regions, and nucleus--nucleus interactions.
The DPMJET model, at the time of investigations, was not available in
CORSIKA in a stable version.
In addition, SIBYLL was used, a minijet model that is widely used in EAS 
calculations, especially as the hadronic interaction model in the MOCCA code.
A sample of 2000 proton and iron--induced showers were simulated with SIBYLL 
and 7000 p and Fe events with QGSJET. 
With VENUS 2000 showers were generated, each for p, He, O, Si and 
Fe primaries. The showers were distributed in the energy range of 
0.1\, PeV up to 31.6\, PeV according to a power law with a
differential index of -2.7 and were equally spread in the interval of $15^o$ to
$20^o$ zenith angle. In addition, the changing of the index to -3.1 at the 
{\it knee} position, which is assumed to be at 5~PeV, was taken into account.
The shower axes were spread uniformly over the calorimeter surface extended 
by 2~m beyond its boundary. 

In order to determine the signals in the 
individual detectors, all secondary particles at ground level are passed 
through a detector simulation program using the GEANT package \cite{geant}. 
By these means, the instrumental response is taken into account and the 
simulated events are analysed in the same way as the 
experimental data, an important aspect to avoid systematic biases by 
pattern recognition and reconstruction algorithms.

\section{Shower size determination}
The data evaluation proceeds via three levels. In a first step 
the shower core and its direction of incidence are reconstructed and,
using the single muon calibration of the array detectors, their energy 
deposits are converted into numbers of particles. In the next stage, 
iterative corrections for electromagnetic 
punch--through in the muon detectors and muonic energy deposits in 
the electron detectors are applied. The particle densities are 
fitted with a likelihood function to the Nishimura Kamata Greisen 
(NKG) formula \cite{kamata}. A radius parameter of 89~m and 420~m is used
for electrons and muons, respectively. Because of limited statistics, the
radial slope parameter (age) is fixed for the muons.
The radius parameters deviate from the parameters originally proposed, 
but have been found to yield the best agreement with the data.
The muon fit extends from 40~m to 200~m, the lower cut being imposed 
by the strong hadronic
and electromagnetic punch--through near the shower centre. The upper 
boundary reflects the geometrical acceptance. In a final step, the 
muon fit function is used to correct the electron numbers and vice 
versa.

\begin{figure}\centering
   \epsfig{file=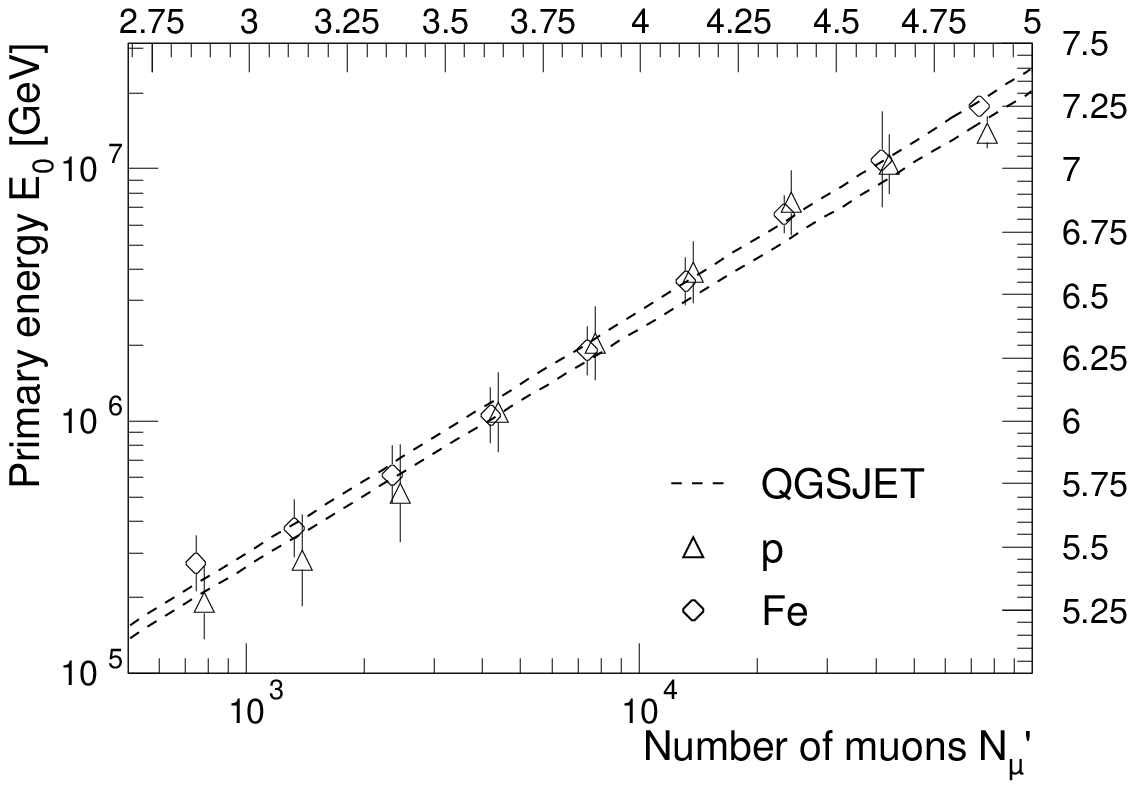,width=0.48\textwidth}\hskip2mm%
   \epsfig{file=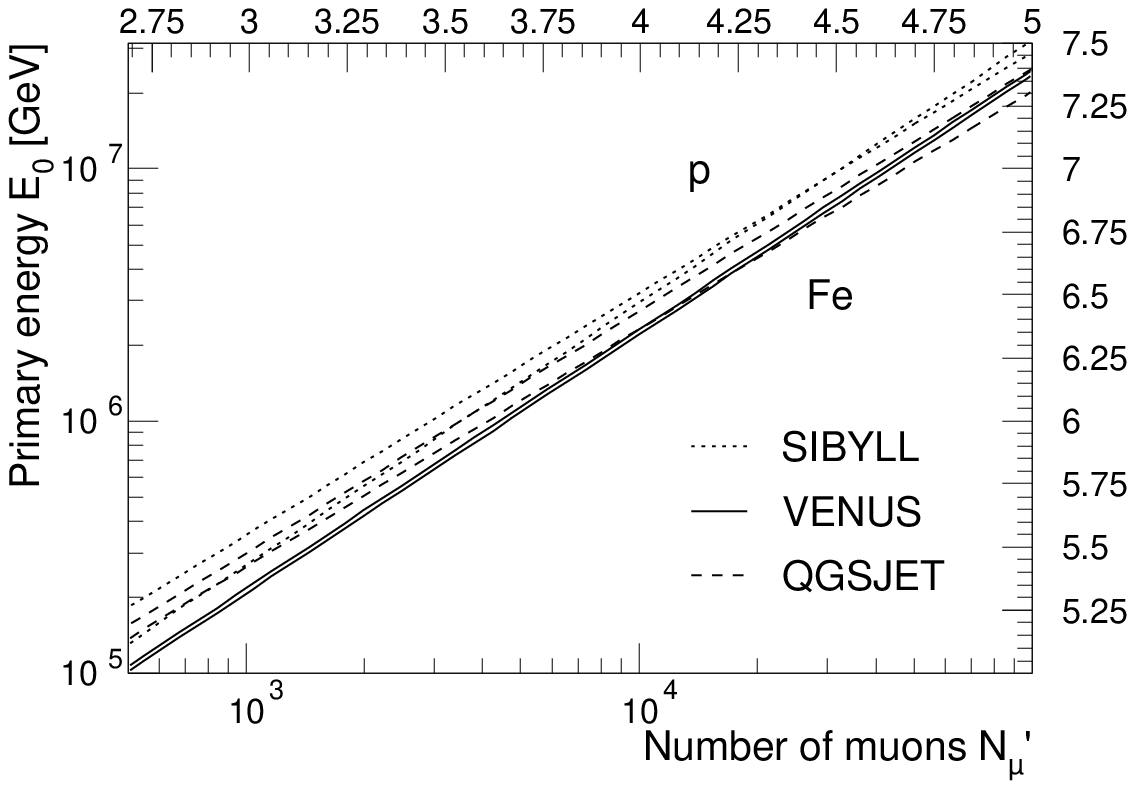,width=0.48\textwidth}%
   \caption{ Primary energy as determined by simulations from 
   the truncated muon number $N'_\mu$ using the interaction models 
   indicated. The vertical bars indicate the rms widths of primary energy 
   distribution for fixed number of muons.}
   \label{eeichnmy}
\end{figure}

The electromagnetic and muonic sizes $N_{e}$ and $N_{\mu}$ are 
obtained by integrating the final NKG fit functions. For the 
muons alternatively, integration within the range of the fit 
results in a truncated muon number $N'_\mu$. This observable has 
the advantage of being free of systematical errors caused by the 
extrapolation outside the experimental acceptance. As 
demonstrated in Fig.~\ref{eeichnmy}, it yields a good estimate
of the primary energy irrespective of primary mass. To a certain extent, 
it is an integral variable indicating the sum of particles produced in the 
atmosphere independently of longitudinal cascade development. 
%It has been checked that the particle numbers are correctly evaluated
%throughout the acceptance range, only at the highest primary energy of
%100~PeV $N'_\mu$ is overestimated by about 8\% due to residual
%punch--through.
In the lefthand graph, the simulated values for the QGSJET model are plotted 
together with fitted straight lines. 
They show that in the $N'_\mu$ range given, the primary energy is 
proportional to the muon number $E_0 \propto N_\mu'^{0.98}$ with an error
in the exponent of 0.06. This holds for the selected showers hitting 
the central detector with their axes. (For all showers falling 
into the area of the array a slightly higher  coefficient of 1.10 is found.)

It has been checked that the particle numbers are evaluated correctly up to 
values of $\lg N'_\mu=5$. At the highest energy of 100~PeV simulations 
indicate that $N'_\mu$ is overestimated by about 10\%. By studying $N_\mu$ 
sizes at this energy experimentally, irregularities in the muon size 
distribution may  indicate an overestimation of 20\%. How well different 
models agree among each other is shown on the righthand part of 
Fig.~\ref{eeichnmy}, where the corresponding fitted lines are presented. 
It is seen that the 
SIBYLL model lies above the two others. In other words it generates 
fewer muons with consequences that will be discussed below.  It is 
this truncated muon number $N'_\mu$ which we shall use throughout this 
article to classify events according to the muon number, that 
means approximately according to the primary energy.

The accuracy of the reconstructed shower sizes is estimated to be 5\% 
for $N_{e}$ and 10\% for $N'_\mu$ around the {\it knee} position.

\section{Hadron reconstruction}
The raw data of the central detector are passed through a pattern 
recognition program which traces a particle in the detector
and reconstructs its position, energy and incident angle. Two algorithms 
exist. One of them is optimized to reconstruct unaccompanied hadrons 
and to determine their energy and angle with best resolution. The 
second is trained to resolve as many hadrons as possible in a shower core 
and to reconstruct their proper energies and angles of 
incidence. This algorithm has been used for the analyses 
presented in the following. Grosso modo, the pattern recognition proceeds 
as follows: Clusters of energy are searched 
to line up and to form a track, from which roughly an angle of
incidence can be inferred. Then in the lower layers patterns of
cascades are looked for since these penetrating and late 
developing cascades can be reconstructed most easily. 
Going upwards in the 
calorimeter, clusters are formed from the remaining energy and 
lined up to showers according to the direction already found. The 
uppermost layer is not used for hadron energy determination to evade 
hadron signals, which are too much distorted by the electromagnetic component,
nor is the trigger layer used because of its limited dynamical range.

Due to a fine lateral segmentation of 25~cm, the minimal distance 
to separate two equal--energy hadrons with 50\% probability 
amounts to 40~cm. This causes the reconstructed hadron density to flatten 
off at about 1.5~hadrons/m$^2$. The reconstruction efficiency with respect
to the hadron energy is presented in Fig.~\ref{effi}. 
At 50~GeV an efficiency 
of 70\% is obtained. This energy is taken as threshold in most 
of the analyses in the following, if not mentioned otherwise. 
We present the values on a logarithmic scale in order to 
demonstrate how often high--energy radiating muons can mimic
a hadron. Their reconstructed hadronic energy, however, is much 
lower, typically by a factor of 10. The fraction of non--identified
hadrons above 100\,GeV typically amounts to 5\%. This value 
holds for  a 1 PeV shower hitting the calorimeter at its centre and rises
to 30\% at 10 PeV. This effect is taken into account automatically, because
in the simulation it appears as the same token.
 
\section{Event selection}
About $10^8$ events were recorded from October 1996 to August 1998. 
In $6\times10^6$ events, at least one hadron was reconstructed. Events 
accepted for the present analysis have to fulfill 
the following requirements: More than two hadrons are
reconstructed, the zenith angle of the shower is less 
than $30^{\circ}$ and the core, as determined by the array stations, hits 
the calorimeter or lies within 1.5 m distance outside its boundary. For 
shower sizes corresponding to energies of more than about 1 PeV, the 
core can also be determined in the first calorimeter layer by the 
electromagnetic punch--through. The fine sampling of the
ionisation chambers yields 0.5~m spatial resolution for the core
position.
For events with 
such a precise core position it has to lie within the calorimeter at 1 m
distance from its boundary. After all cuts, 40\,000 events were left
for the final analysis.

For non--centric showers, hadronic observables like the number of hadrons 
have been corrected for the missing calorimeter surface by requiring 
rotational symmetry. On the other hand, some
variables are used, for which such a correction is not obvious, e.g. the 
{\it minimum--spanning--tree}, see Section 8.4. In these cases, only a
square 
of $8\times8~\mbox{m}^2$ of the calorimeter with the shower core in its 
centre is used and the rest of the calorimeter information neglected. 
This treatment ensures that all events are analysed on the same
footing.

\section{Tests at large distances}
\begin{figure}\centering
 \begin{minipage}[t]{0.5\textwidth}
   \epsfig{file=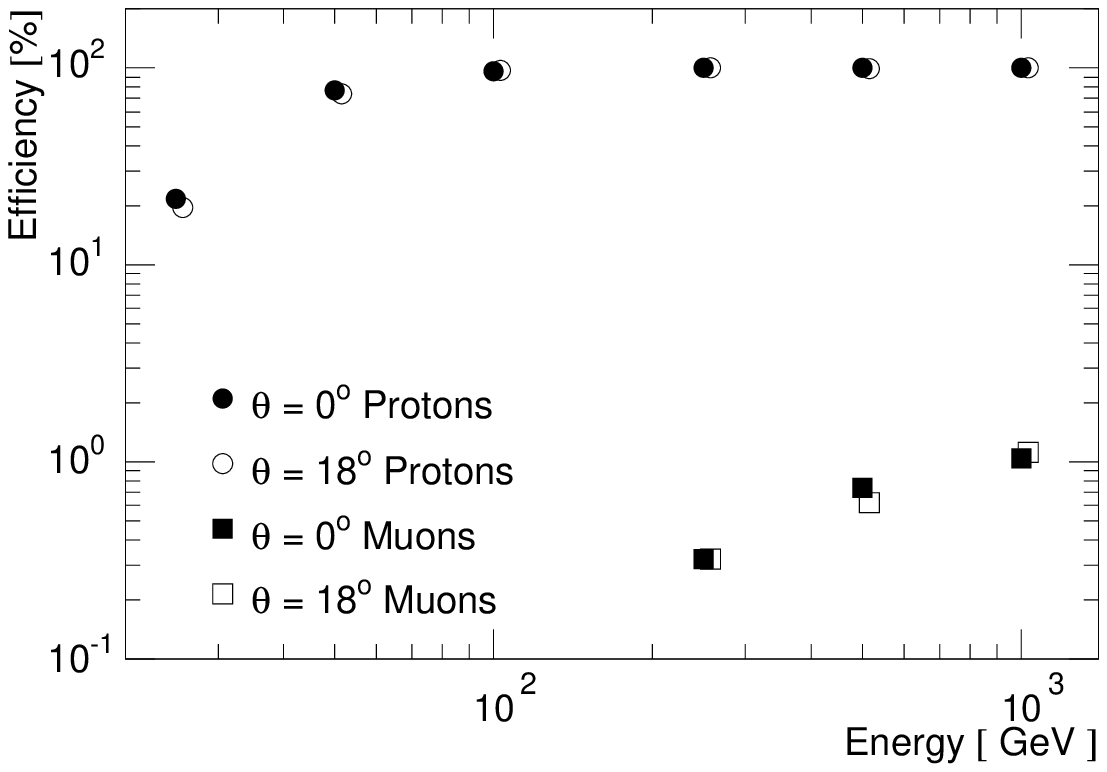,width=\textwidth}
   \caption{Reconstruction efficiency for hadrons for two different
   zenith angles. The square symbols represent the probability of
   radiating muons misidentified as hadrons.}  
   \label{effi}
 \end{minipage}%
 \begin{minipage}[t]{0.5\textwidth}
  \epsfig{file=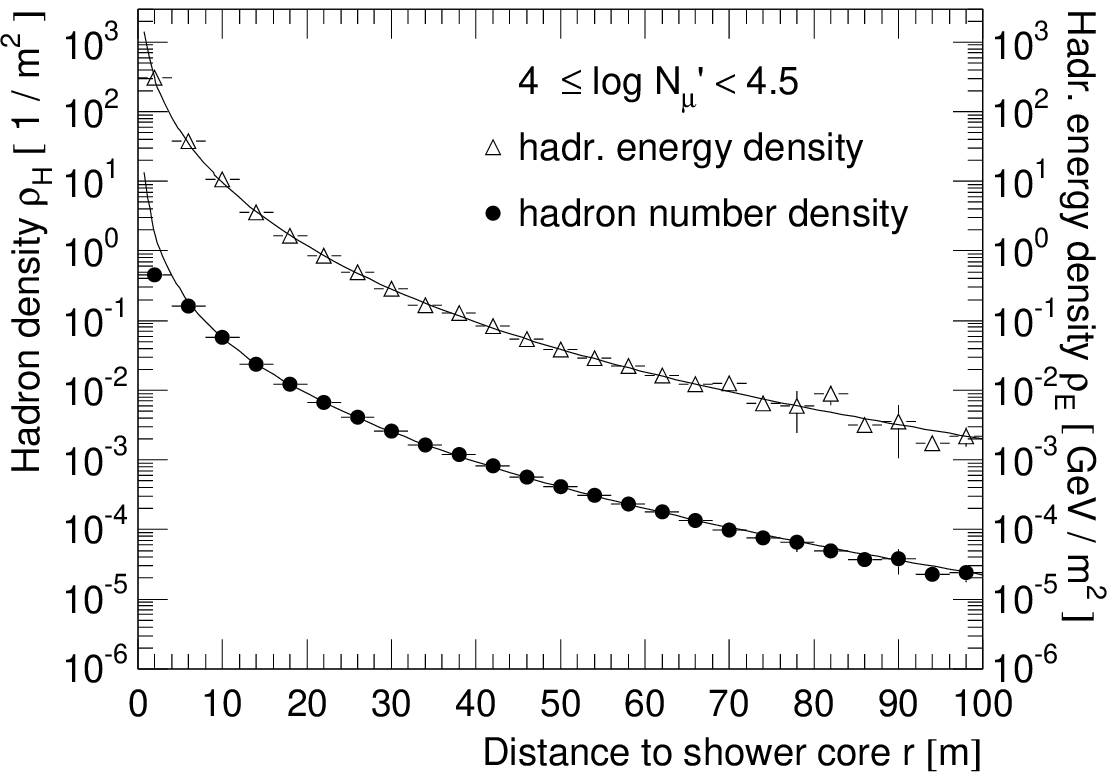,width=\textwidth}%
  \caption{Density of hadrons (left scale) and of hadronic
  energy (right scale) for showers of
  truncated muon numbers as indicated corresponding to primary energies 
  of 3--10~PeV.  The curves represent fits of the NKG formula to the data.}
  \label{rho2}
 \end{minipage}%
\end{figure}

\begin{figure}\centering
 \epsfig{file=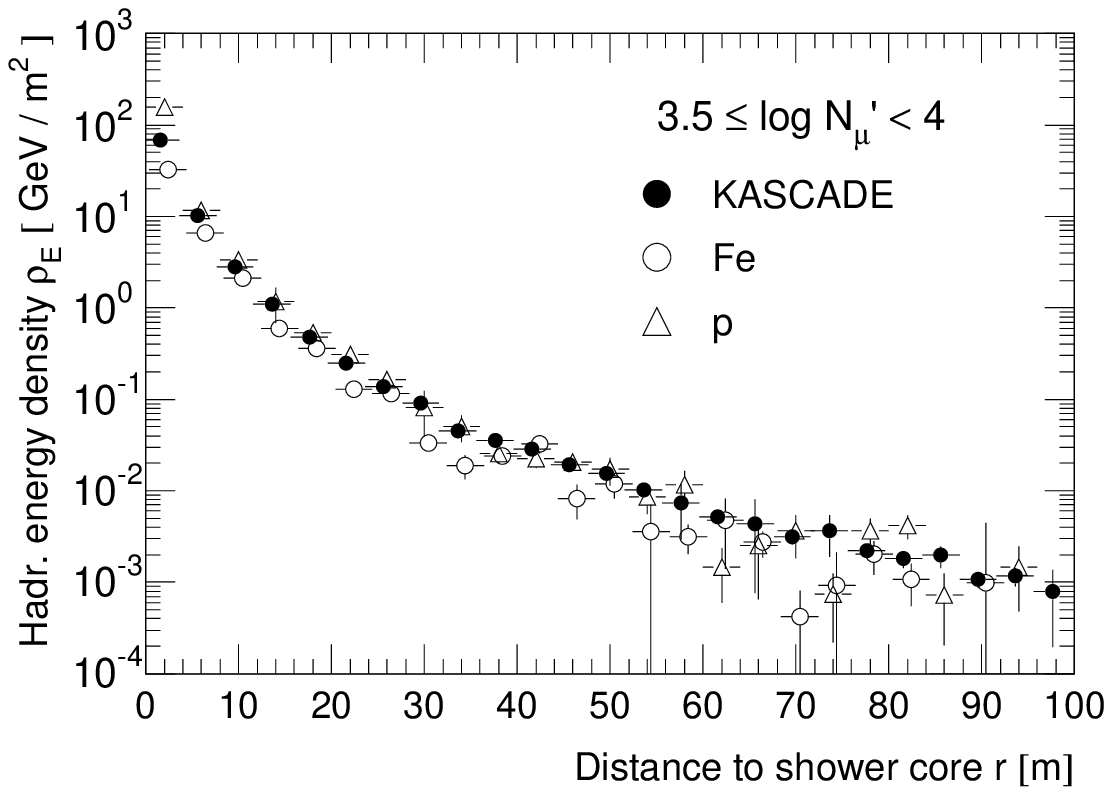,width=0.5\textwidth}%
 \epsfig{file=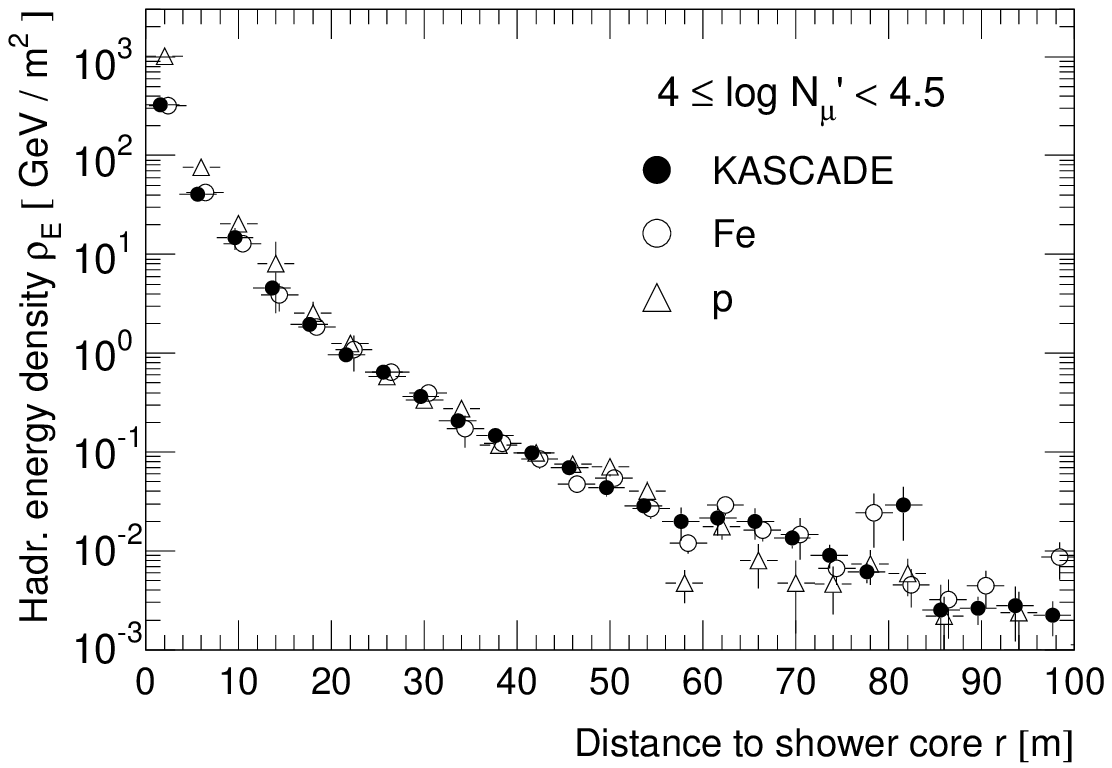,width=0.5\textwidth}%
 \caption{Density of hadronic energy (filled circles) vs. core
 distance for two intervals of primary energy. The indicated muon numbers
 correspond to $1~\mbox{PeV}\le E_0 < 3~\mbox{PeV}$ and 
 $3~\mbox{PeV}\le E_0<10~\mbox{PeV}$. 
 The CORSIKA simulations (open symbols) represent
 primary proton and iron nuclei with the QGSJET model.}
 \label{rho}
\end{figure}

\begin{figure}
 \epsfig{file=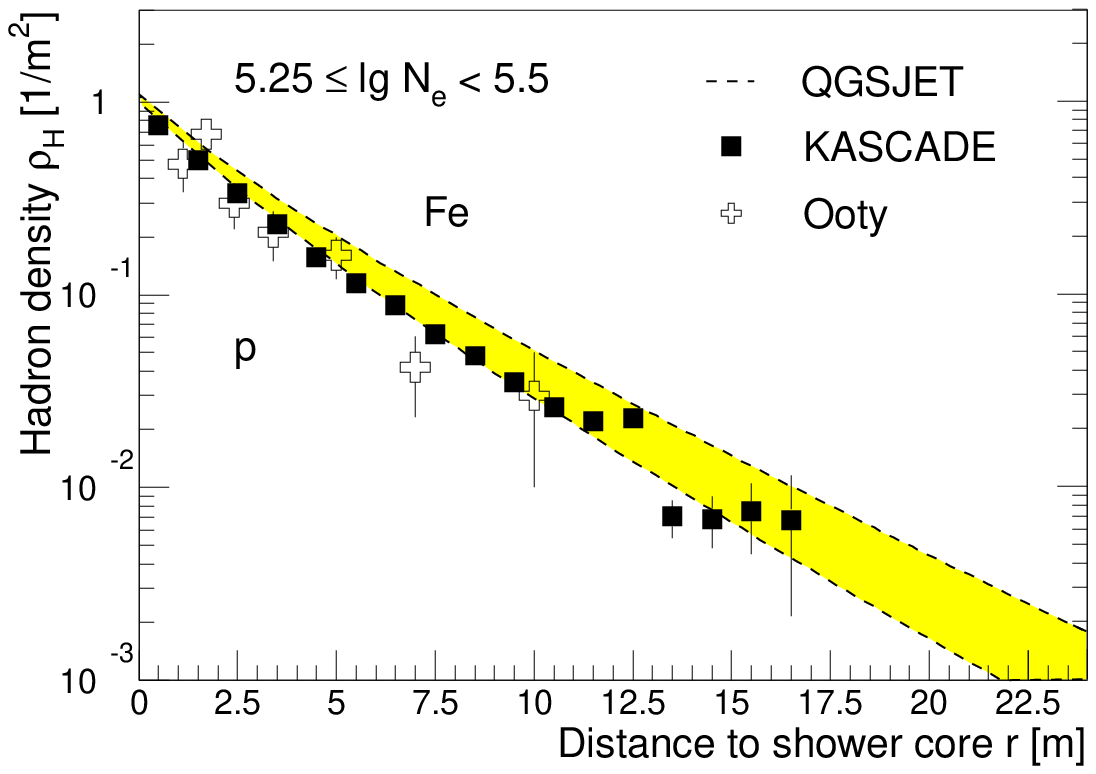,width=0.5\textwidth}%
 \epsfig{file=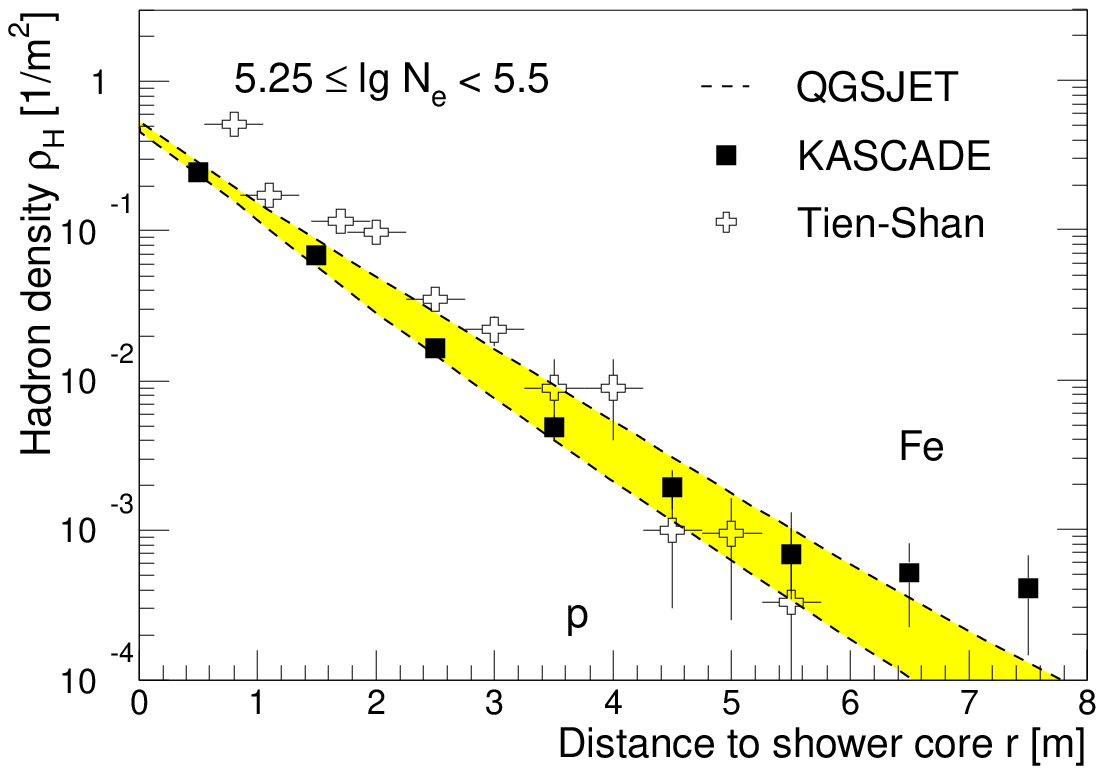,width=0.5\textwidth}
 \caption{Lateral hadron density for electromagnetic shower sizes of $5.25 \le
 \lg N_{e} < 5.5$. Thresholds for hadron detection are 50~GeV (left), 
 and 1~TeV (right). The dashed lines represent CORSIKA simulations with
 the QGSJET model for primary proton and iron nuclei using an 
 exponential, see text.}
 \label{ootytien}
\end{figure}

\begin{figure}\centering
 \epsfig{file=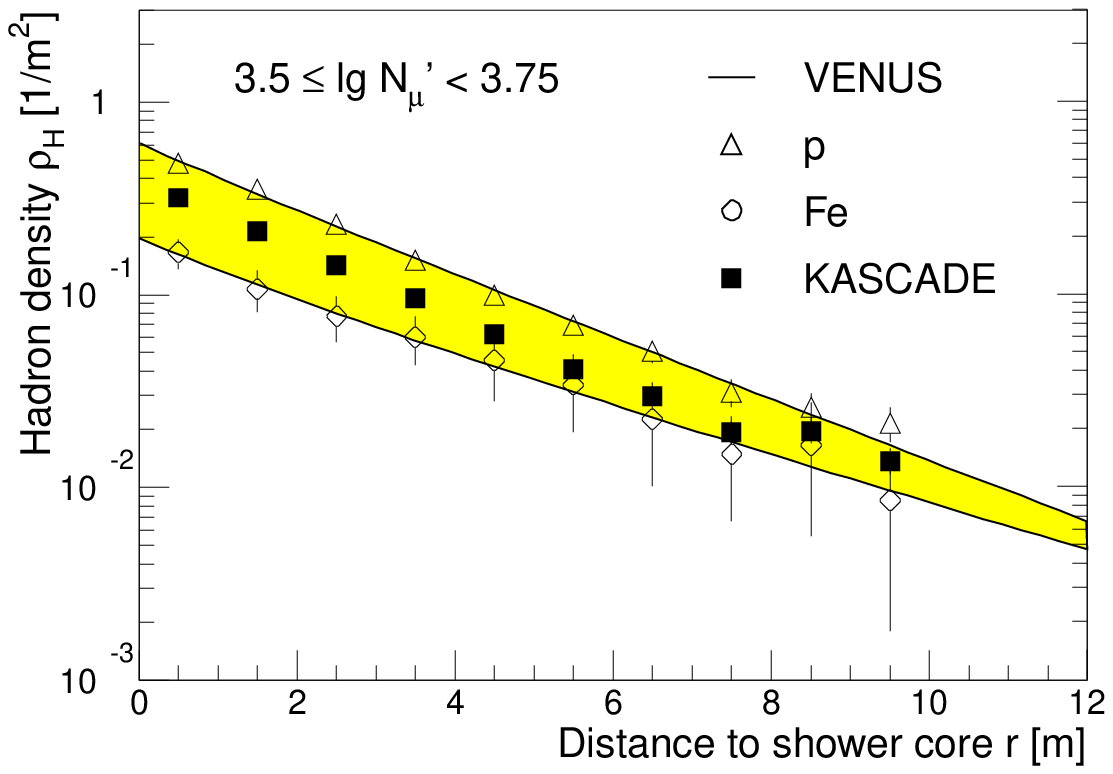,width=0.5\textwidth}%
 \epsfig{file=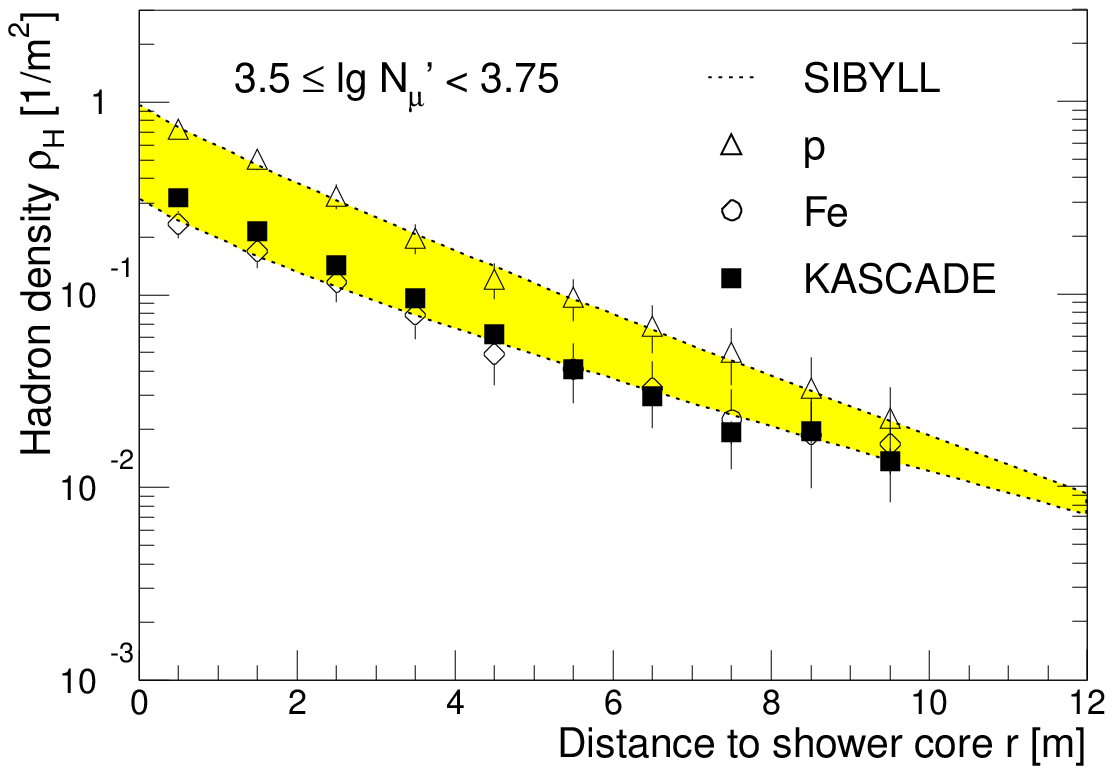,width=0.5\textwidth}%
 \caption{The lateral hadron density for muonic shower sizes
 corresponding to a mean primary energy of 1.2 PeV. Threshold for
 hadron detection is 50~GeV. The data are compared with simulations using
 VENUS  (left) and SIBYLL (right), the curves represent fits according to
 a modified exponential, see text.}
 \label{mlat}
\end{figure}

Studying hadron distributions at large core distances
checks mainly the overall performance of the shower simulation 
program CORSIKA. In the regions far away from the shower axis of an EAS,
the Monte Carlo calculations can be verified with respect to the
transport
of particles, their decay characteristics, etc. If the hadrons are well
described it
signifies that the shower propagation is treated properly. In these
outer 
regions, where lower hadron energies and larger scattering angles dominate, 
the underlying physics is sufficiently well known from accelerator
experiments, and the code in itself can be tested. 

As an example of such a test, the hadron lateral distribution is
presented in Fig.~\ref{rho2} for $N'_\mu$ sizes corresponding to the primary 
energy interval around and above the {\it knee}: 
$3~\mbox{PeV}\le E_0<10~\mbox{PeV}$. 
The distributions of the number of hadrons and of the hadronic 
energy are given. In the very centre of the former, a saturation 
as mentioned in chapter 5 can be noticed. Several 
functions have been tried to fit the data points, among others 
exponentials as suggested by Kempa \cite{kempa}. However, by far the best fit 
was obtained when applying the NKG formula represented by the 
curves shown in the graph. This finding is not particularly 
surprising because hadrons of an energy of approximately 100~GeV \cite{kempa},
when passing through the atmosphere, generate the electromagnetic component, 
and the NKG formula has been derived for electromagnetic cascades. 
In addition, multiple scattering of electrons 
determining the Moli\`ere radius resembles the scattering character 
of hadrons with a mean transverse momentum of 400~MeV/c 
irrespective of their energy. Replacing the mean multiple scattering 
by the latter and the radiation length by the interaction length 
one arrives at a radius $R_H$ of about 10~m. This 
value we expect to take the place of the Moli\`ere radius in the 
NKG formula for electron measurements. Indeed, values of this order are 
found experimentally.

Lateral hadron distributions compared with CORSIKA simulations 
are shown in Fig.~\ref{rho} for primary energies below and above the 
{\it knee}. In the diagrams the hadronic energy density is plotted for 
muon numbers corresponding to the primary energy intervals of 
$1~\mbox{PeV}\le E_0<3~\mbox{PeV}$ and $3~\mbox{PeV}\le E_0<10~\mbox{PeV}$. 
The data points are compared to primary proton and iron simulations 
applying the QGSJET model. These two extreme assumptions about the masses 
result in nearly identical hadron densities and the measured data coincide 
with the simulations, thereby verifying the calculations. Similar good
agreement is found for the VENUS and SIBYLL 
models. Simulations and data agree well up to 100~m distance 
from the core. Only in the very inner region of 10~m, the simulations 
yield deviating hadron densities for different primary masses. 
Nevertheless, the measurements
here lie well in between the two extreme primary compositions of pure protons 
or pure iron nuclei.

\section{Tests at shower core}
\subsection{Hadron lateral distribution}
To begin with, the lateral distributions are compared with values 
published in the literature. Hadron 
distributions in the core of EAS have been measured at Ooty by Vatcha 
and Sreekantan \cite{vatcha} and at Tien Shan by 
Danilova et al. \cite{danilova}. Results of
earlier experiments have been examined and discussed 
by Sreekantan et al. \cite{sreekantan}. 
In the experiments different techniques for hadron 
detection have been applied: a cloud chamber at Ooty, long gaseous 
ionisation tubes at Tien Shan and liquid ionisation chambers in the 
present experiment. Therefore, it is of interest to compare the 
respective results.

The experiments were performed at different altitudes, and a 
priori they are expected to deliver deviating results. However, when 
compared at the same electromagnetic shower size, hadron distributions should 
be similar because electrons and hadrons, the latter of about 100 GeV, 
are closely related to each other in an EAS when the shower passes through the 
atmosphere. A sort of equilibrium turns up as has been pointed out by 
Kempa \cite{kempa}. Indeed, Fig.~\ref{ootytien} demonstrates for electron 
numbers $5.25 \leq \lg N_{e} < 5.5$ that the lateral hadron distributions 
agree reasonably well. In particular, the measurements of the Ooty 
group at an atmospheric depth of $800~\mbox{g}/\mbox{cm}^2$ coincide with the 
present findings. The grey shaded band represents CORSIKA 
simulations using the hadronic interaction model QGSJET, the lower 
curve representing primary protons and the upper curve primary iron 
nuclei. The curves are fits to the simulated density of hadrons
according to $\rho_H(r)\propto exp~(-(\frac{r}{r_0})^\kappa)$ with
values for $\kappa$ found to be between 0.7 and 0.9. The data lie well 
between these two boundaries. The graph on the righthand side represents 
hadron densities with a threshold of 1 TeV. Bearing 
this high threshold in mind, the similarity in both distributions, Tien 
Shan at $690~\mbox{g}/\mbox{cm}^2$ and KASCADE at sea--level, 
is astonishing. In 
conclusion, it can be stated that hadron densities, despite of being measured 
with different techniques, agree reasonably well among different
experiments.

When classifying hadron distributions according to muonic shower sizes, 
differences among the interaction models emerge. This becomes 
apparent in Fig.~\ref{mlat}, where the central density is plotted 
for truncated muon numbers which correspond to a mean energy of about 1.2 PeV.
On the left graph, the VENUS calculations enclose the data points leaving 
the elemental composition
to be somewhere between pure proton or pure iron primaries. On the 
right graph, the measured data points follow the lower boundary of the
SIBYLL calculations, suggesting 
that all primaries are iron nuclei, at this energy obviously an 
unprobable result. 

The lateral distribution demonstrates, and other observables in a similar 
manner as reported previously \cite{jrh}, that the SIBYLL code generates too low muon numbers thereby entailing
a comparison at a different estimate of the primary energy. 
A hint has 
already been observed in Fig.~\ref{eeichnmy} where the SIBYLL lines lie above 
those of QGSJET and VENUS.  When hadronic observables are classified 
according to electromagnetic shower sizes, the disagreement vanishes as will be
discussed in the following. 

\subsection{Hadron energy distribution}
\begin{figure}\centering
 \begin{minipage}[t]{0.5\textwidth}%
  \epsfig{file=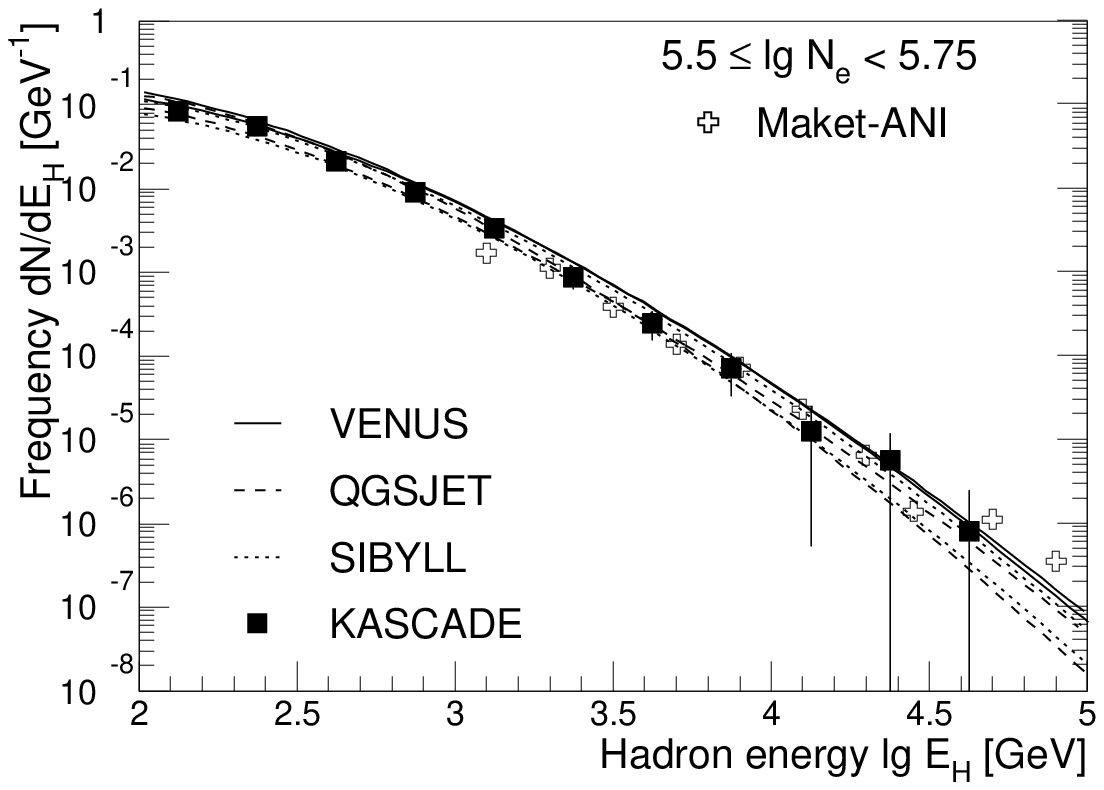,width=\textwidth}%
  \caption{Hadron energy distribution for fixed electron number
  $N_{e}$  corresponding approximately to 6 PeV primary energy. The
  lines represent CORSIKA simulations with three interaction models, the
  lower curves for primary iron, the upper for protons.}
  \label{mespec}
 \end{minipage}%
 \begin{minipage}[t]{0.5\textwidth}%
  \epsfig{file=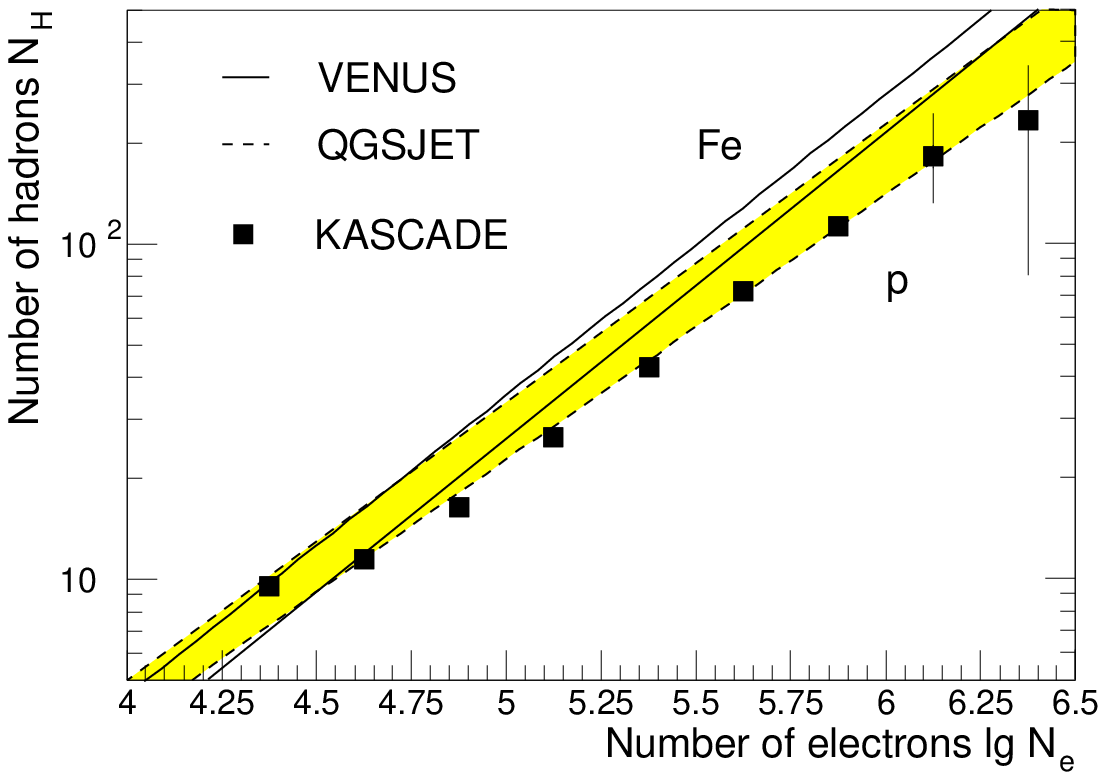,width=\textwidth}%
  \caption{Hadronic shower size vs. electromagnetic shower size. 
  Experimental values are
  compared with simulations using VENUS (full lines) and QGSJET (shaded area),
  both for primary protons and iron nuclei. The experimental error 
  bars are rms--values.}
  \label{nintne}
 \end{minipage}%
\end{figure}

The energy distribution of hadrons is shown in Fig.~\ref{mespec} for a 
fixed electromagnetic shower size. Plotted is the number of hadrons in an 
area of $8\times8~\mbox{m}^2$ around the shower core. As 
already mentioned, in this way all showers are treated in the same 
manner, independent of their point of incidence. 
To avoid a systematic bias the
loss in statistics has to be accepted. The number of showers reduces to about 
5000. The shower size bin of $5.5 
\leq \lg N_{e} < 5.75$ corresponds approximately to a mean primary
energy of 6~PeV.
The lines represent fits to the simulations according to
$exp~(-(\frac{\lg E_H-a}{b})^c)$. Usually in the literature
$c=1$ is assumed, however, the present data, due to their large dynamical
range, yield values for $c$ from 1.3 to 1.6.
As can be inferred from the graph, all three interaction models 
reproduce the measured data reasonably well, elucidating the fact 
that electrons closely follow the hadrons in EAS propagation. But if 
the same data are classified corresponding to the muon number, again 
SIBYLL seems to generate too many hadrons and thereby mimics a  
primary composition of pure iron nuclei. 
For this reason SIBYLL will not be utilized any further.
In the figure also the energy spectrum is
plotted as measured with the Maket--ANI calorimeter by Ter--Antonian et 
al.\ \cite{ter}. As already mentioned above, distributions are expected to 
coincide when taken at the same electron number even if they have been 
measured at different altitudes. In the present case the data have been 
taken at sea level and at $700~\mbox{g}/\mbox{cm}^2$ on Mount Aragats. 
The energy distributions, indeed, agree rather well with each other 
indicating that in both data sets the patterns of hadrons are well 
recognized and the energies correctly determined. 

It was seen that SIBYLL encounters difficulties when the data are classified 
according to muonic shower sizes. The model VENUS, on the other hand, cannot 
reproduce hadronic observables convincingly well when they are binned into
electron number intervals. An example is given in Fig.~\ref{nintne}. 
It shows the number 
of hadrons, i.e. the hadronic shower size $N_h$, 
as a function of the electromagnetic shower size 
$N_e$. The experimental points match well to the primary proton line as 
expected from QGSJET predictions. This phenomenon is easily understood 
by the steeply falling flux 
spectrum and the fact that primary protons induce larger electromagnetic
sizes at observation level than heavy primaries. 
Hence, when grouping in $N_{e}$ bins ,
showers from primary protons will be enriched and we expect to have 
predominantly proton showers in our sample.
This fact reduces any ambiguities in the results due to the absence of
direct information on primary composition.
Concerning the VENUS model the predicted hadron numbers are too high and
the two lines which mark the region between primary
protons and iron nuclei cannot explain the data. The point at the lowest
shower size is still influenced by the trigger efficiency of the array 
counters.

\subsection{Hadron energy fraction}
\begin{figure}\centering
 \epsfig{file=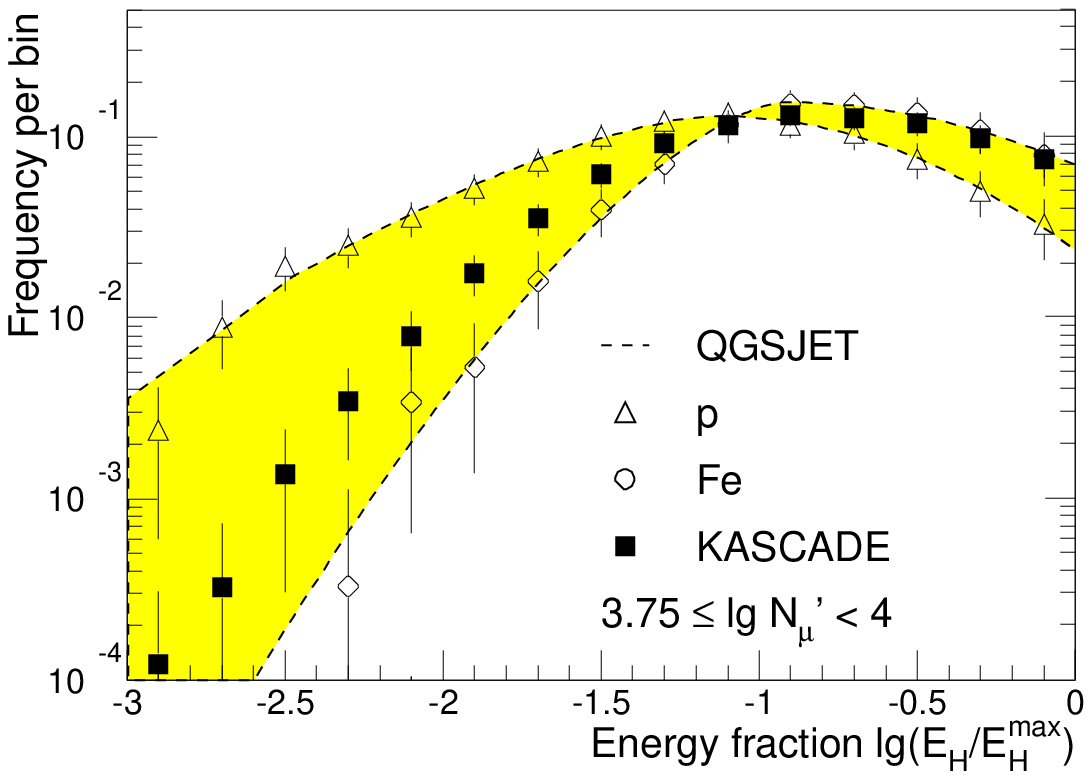,width=0.5\textwidth}%
 \epsfig{file=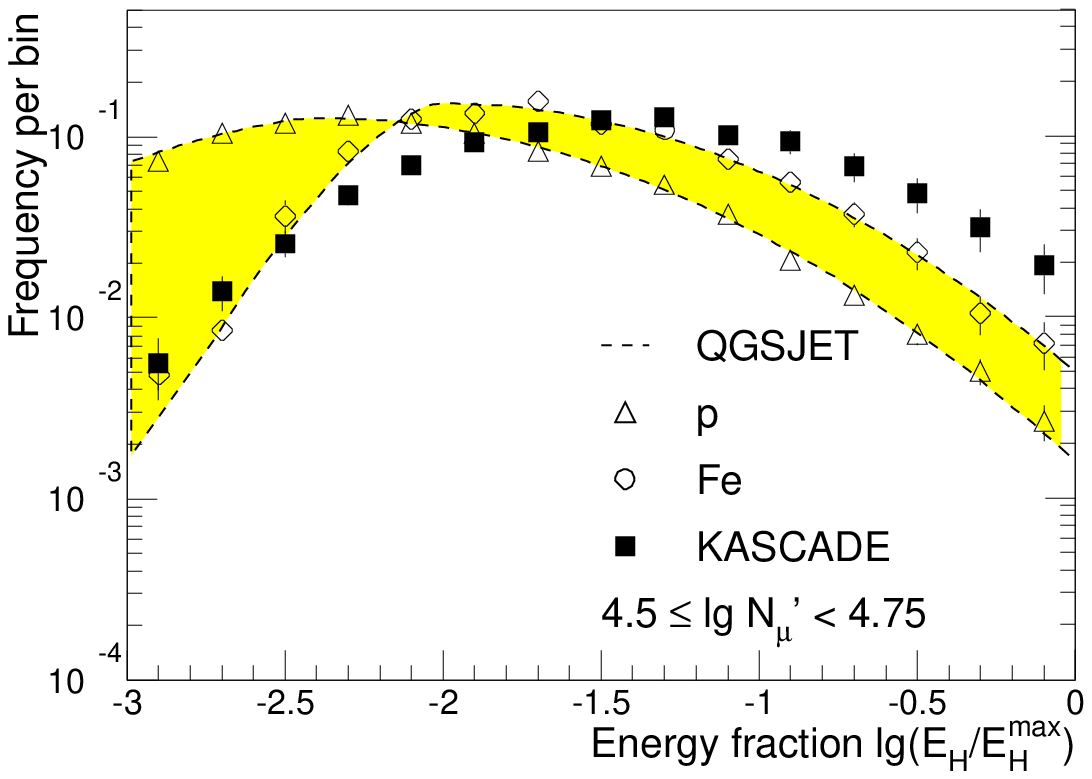,width=0.5\textwidth}%
 \caption{The energy fraction of all hadrons vs. the most energetic
 hadron in a shower. The data are compared to simulations using the
 QGSJET model for primary protons (p) and iron nuclei (Fe). Shaded is
 the physically meaningful region as obtained from the simulations.
 Primary energies correspond to 2 PeV (left) and 12 PeV (right).}
 \label{mefracl}
\end{figure}

A suitable test of the interaction models consists in investigating the 
granular structure of the hadronic core concerning spatial as well as
energy distributions. As variables we have chosen the energy fraction of 
hadrons and the distances in the {\it minimum--spanning--tree} 
between them. Both will be dealt with in the following sections. 

For each hadron its energy fraction with respect to the 
most energetic hadron in that particular shower is calculated. For 
primary protons, the leading particle effect is expected to produce 
one particularly energetic hadron accompanied by hadrons with a broad 
distribution of lower energies. Hence, we presume to find a rather large
dispersion of hadronic energies for primary protons, whereas for primary
iron nuclei the hadron energies should be more equally distributed.
The simulated distributions, indeed, 
confirm this expectation as is shown in Fig.~\ref{mefracl}. 
The lines --- to guide the eye --- represent fits to the simulations using two
modified exponentials as in the preceding section, which are connected to
each other at the maximum.
On the lefthand graph, the data seem to corroborate the simulations. 
They are shown for a muon number range corresponding to a primary energy of 
approximately 2 PeV, i.e., below the  {\it knee} position. On the righthand
side, the results are shown for an interval above the {\it knee} for 
muonic shower sizes corresponding to a primary energy of 12 PeV. The
reader observes 
that the data cannot be explained by the simulations, neither by primary
protons nor by iron nuclei. On a logarithmic 
scale the data exhibit a symmetric distribution around the value 
$\lg (E_H/E_H^{max})\cong-1.5$, even more symmetric than  
would be expected
for a pure iron composition. In particular, 
energetic hadrons resulting from the leading particle effect 
seem to be missing. They would shift the distribution to smaller values. 
This absence of energetic hadrons in the observations will be confirmed
later when investigating other observables. 

\subsection{Minimum--spanning--tree}
\begin{figure}\centering
 \epsfig{file=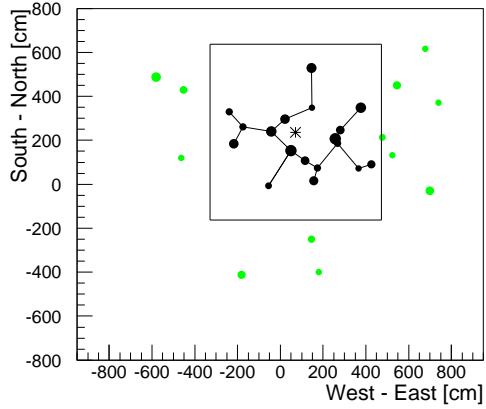,width=0.5\textwidth}%
 \caption{Example of a hadronic core in the calorimeter (top view).
 The square marks the acceptance area of $8\times8~\mbox{m}^2$ around the
 shower centre (star). The energy of each hadron is indicated by the
 area of its point in a logarithmic scale.}
 \label{mst}
\end{figure}

\begin{figure}\centering
 \epsfig{file=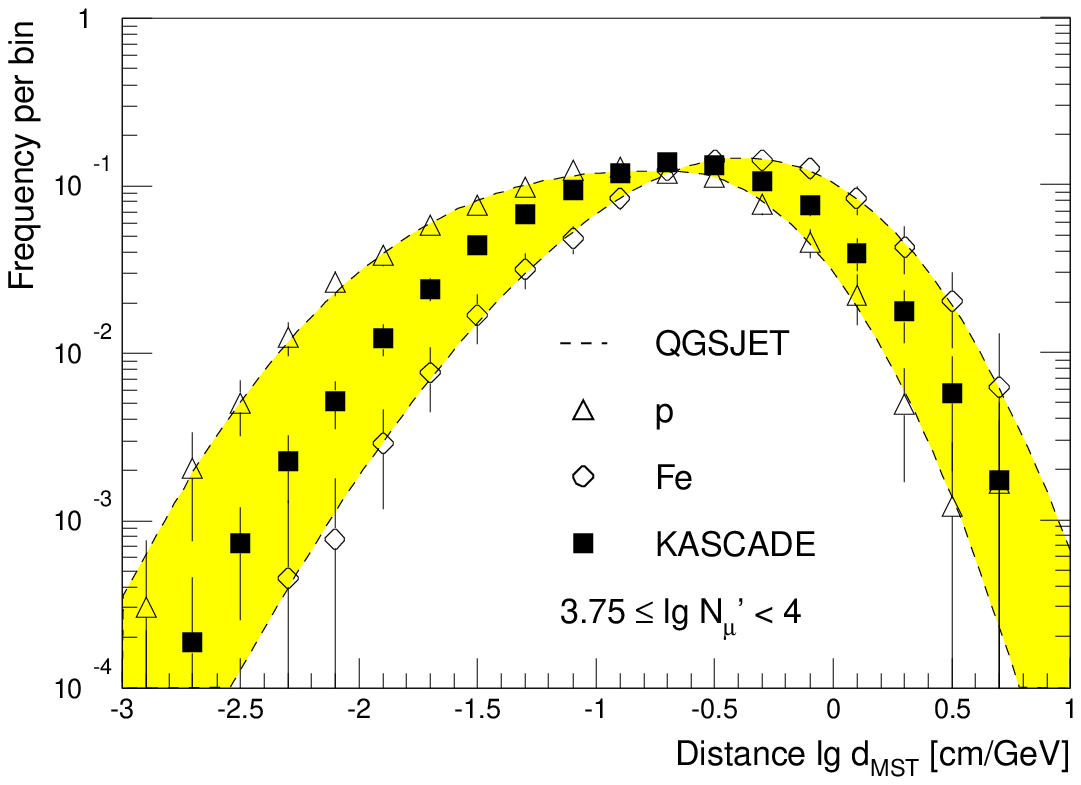,width=0.5\textwidth}%
 \epsfig{file=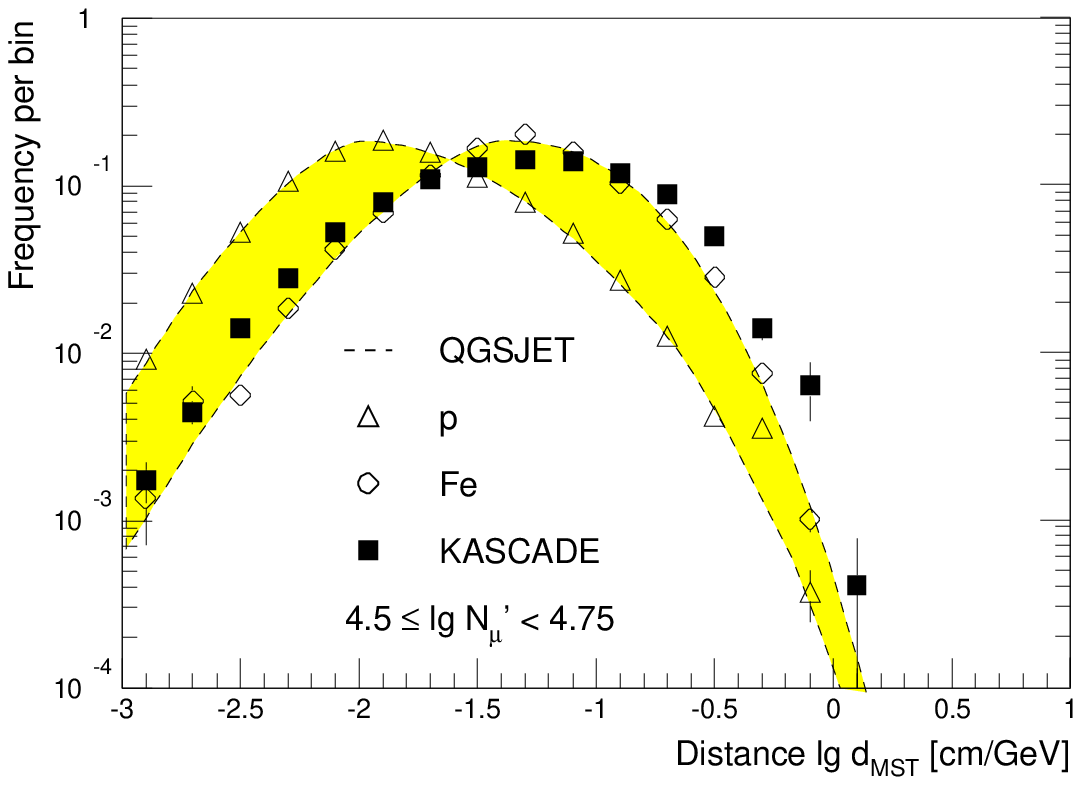,width=0.5\textwidth}%
 \caption{The distances in a minimum--spanning--tree for a muon number
 interval below (left) and above (right) the {\it knee}. The
 measurements
 are compared with simulations using the interaction model QGSJET for
 primary protons (p) and iron nuclei (Fe). The lines are fits to the
 simulations analogous to Fig.~\ref{mefracl}.}
 \label{mmsts}
\end{figure}
When constructing the {\it minimum--spanning--tree} (MST), all hadrons are 
connected to each other in a plane perpendicular to the shower axis. 
The MST is that configuration where the sum of all connections 
weighted by the inverse energy sum of its neighbours has a minimum. 
The $1/E$ weighting has been found to separate iron and proton induced
showers the most.
Fig.~\ref{mst} shows as an example the central shower core of an event. 
Plotted are the points of incidence on the calorimeter. The sizes 
of the points mark the hadron energies on a logarithmic scale. The 
shower centre and the fiducial area of $8\times8~\mbox{m}^2$ around 
it are indicated as well. 
For each event the distribution of distances is formed. Average distributions 
from many events are given in Fig.~\ref{mmsts}.
As in Fig.~\ref{mefracl}, the muonic shower sizes correspond to 
primary energy intervals below and 
above the {\it knee}. It is observed that for the former, the data lie well 
within the bounds of the primary composition but that above the 
{\it knee} the measurements yield results which are not in complete agreement
with the model although they are close to the simulated iron data. The
distributions of Figs.~\ref{mefracl} and \ref{mmsts} have also been calculated
analysing the full calorimeter 
surface and not only the $8 \times 8~\mbox{m}^2$ around the shower centre. No 
remarkable difference could be obtained. 

In both observables -- energy fraction and MST --
the data for higher primary energies cannot be interpreted by the simulations.
Additionally, in the M.C. calculations the {\it knee} in the primary energy 
distribution has been omitted. Again, no remarkable change in the
distributions showed up. In fact, when investigating the distributions as a 
function of muon number, the deviation between M.C. values and the 
measured data develops smoothly with increasing energy.

When regarding the righthand graph in Fig.~\ref{mmsts} the question arises 
whether the interaction model produces too small distances or too energetic 
hadrons or both. In agreement with the observation in Fig.~\ref{mefracl} one
has to conclude that too energetic hadrons are generated as compared to the 
data. Whether in the MSTs also the distances between the hadrons, in other 
words the transverse momenta, are underestimated cannot be decided at the 
moment. Also the number of hadrons plays a role. This issue is under further 
investigations.

\subsection{Hadronic energy in large showers}
\begin{figure}\centering
 \epsfig{file=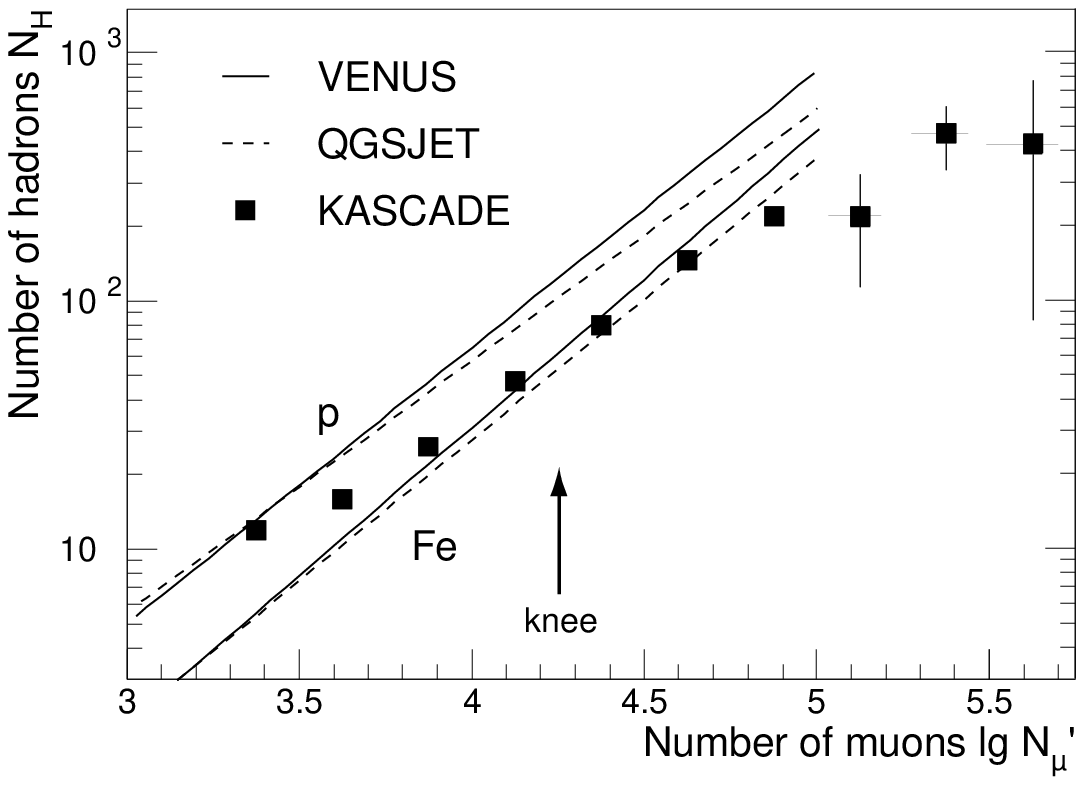,width=0.5\textwidth}%
 \epsfig{file=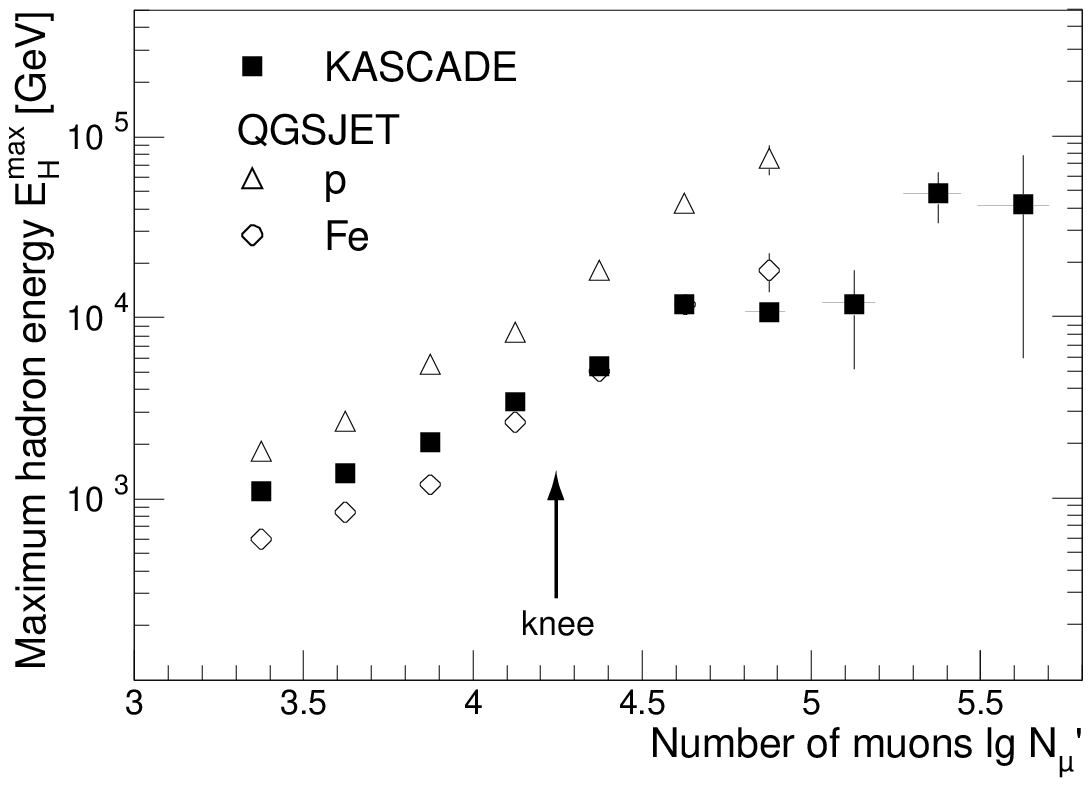,width=0.5\textwidth}
%	 clip=,bbllx=150,bblly=323,bburx=463,bbury=548}%
 \caption{Hadronic shower size vs. muonic shower size (left) 
 and the maximum hadron
 energy in a shower vs. its muonic shower size (right). The lines represent
 simulations with the indicated interaction codes. The upper curves
 represent primary protons, the lower primary iron nuclei.}
 \label{nintnmy}
\end{figure}

\begin{figure}\centering
 \epsfig{file=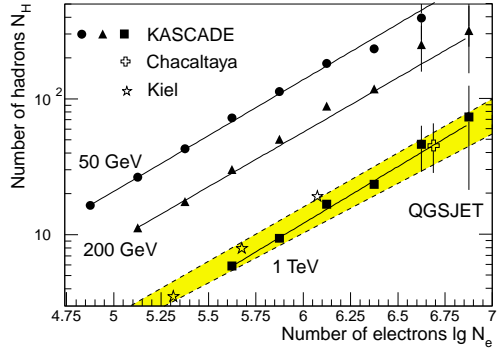,width=0.5\textwidth}%
 \caption{Hadron number at different thresholds, 50 GeV (diamonds),
 200 GeV (triangles), and 1 TeV (squares) vs. electromagnetic shower size. 
 The dashed band
 represents QGSJET simulations for primary protons (lower line) and iron nuclei (upper line).}
 \label{hzahl}
\end{figure}

Deviations between measurement and simulations as in the 
preceding sections are also observed when investigating the hadronic 
energy in large showers. With rising muon numbers $N'_\mu$, the 
experiment reveals an increasing part of missing hadronic energy in 
the shower core. Fig.~\ref{nintnmy} (left) shows 
the number of hadrons versus the muonic shower size.
At muon numbers corresponding to about 5 PeV 
primary energy, the hadron numbers turn out to be 
smaller than predicted for iron
by both interaction models VENUS and QGSJET. Again, one observes that the 
latter model describes the experimental points somewhat better. 
The conclusion that QGSJET reproduces the data best in the PeV region is 
also confirmed by a recent model comparison performed by
Erlykin and Wolfendale \cite{wolfendale}.
The authors classify the models on the basis of consistency checks among 
different observables, e.g.\ the depth of shower maximum $X_{max}$ and the 
$N_\mu / N_e$ ratio.

The righthand graph of Fig.~\ref{nintnmy} presents the maximum hadron 
energy found in showers with the indicated muon number. 
%The curves representing the simulations are fits according to 
%$$E_H^{max}=A\cdot e^{\left(\frac{\lg N'_\mu}{B}\right)^{3.5}}\quad.$$
The open symbols represent the QGSJET simulations, again QGSJET and VENUS
yield similar results.
Measurement and simulation 
disagree also to some extent in this variable at large shower sizes. 
The overestimation of muon numbers mentioned in section 4 cannot account for
the discrepancies. On a logarithmic scale it starts to be noticeable at
$\lg N'_\mu=5.5$ and amounts to $\Delta\lg N'_\mu=0.1$. A shift of this 
size does not ameliorate the situation. The data have been 
checked independently in the reduced fiducial area of $8\times8~\mbox{m}^2$.
But in this analysis, too, the data seem to fake pure iron primaries at 
$\lg N'_\mu=4.3$ and are below that boundary for larger muonic sizes.
Factum is that we do not 
observe the energetic hadrons expected from the M.C. calculations. 
In the energy region 10 to 100~PeV even QGSJET fails to describe the
measurements.

Obviously, the question arises whether these experimentally detected 
effects are artifacts caused, for instance, by saturation effects in the 
calorimeter or by insufficient pattern recognition presumed to be different 
in the simulation than present in 
the experimental data. After all, the high--energy values 
correspond to primary energies of about 100 PeV where 400 hadrons have 
to be reconstructed. At this point, it may be noted again that 
always the experimental and simulated data are compared with each 
other on the detector signal level, hence, a possible hadron
misidentification applies to both data sets. As already pointed out in
section 2, individual hadrons up to 50~TeV 
have been reconstructed and their saturation effects have been examined 
thoroughly.

Some misallocation of energy to individual hadrons might occur, though, 
if lateral distributions of hadrons in the core differ markedly between 
simulations and reality. There may be indication from emulsion 
experiments for this \cite{tamada}. However, from the results shown in 
Fig.~\ref{ootytien} and \ref{mlat} we would not expect any dramatic effect.

Fig.~\ref{hzahl} demonstrates that for large electromagnetic shower sizes, 
the number of hadrons compares well with other experiments as well as with 
CORSIKA simulations. In the diagram the number of hadrons 
above the indicated thresholds is presented with respect to the 
shower size. The values obtained for hadrons above 1~TeV can be 
related to two other experiments performed at Kiel by Fritze et al. 
\cite{fritze} and on the Chacaltaya by C. Aguirre et al. \cite{aguierre}. 
It is observed that 
up to shower sizes which correspond to about 20 PeV for primary 
protons all high--energy hadrons are reconstructed, i.e., more than 
70~TeV energy are found in the calorimeter. When compared to QGSJET 
simulations, the data lie within the physical boundaries as shown for 
the 1~TeV line. On closer inspection the data indicate an increase of the mean
mass with rising energy. 
Also in Fig. \ref{nintne} it has been seen that the hadron numbers are
well reproduced by QGSJET up to the highest electromagnetic shower
sizes.
In conclusion, it can be stated that the 
hadron component compares well between different experiments and with 
M.C. calculations when classified according to electromagnetic shower sizes, 
and that the deviations observed in muon number binning 
cannot be accounted for by experimental imperfections. 

\section{Conclusion and outlook}
Three interaction models have been tested by examining the hadronic 
cores of large EAS. It turned out that QGSJET reproduces the data 
best, but at large muonic shower sizes, i.e.\ at energies above the {\it knee}
even this model fails to 
reproduce certain observables. Most importantly, the model predicts more 
hadrons than are observed experimentally.

The current investigation is a first 
approach with a first data sample of the KASCADE experiment. 
Better statistics, both in the data and in the Monte Carlo 
calculations, are imperative, especially above the {\it knee} in the 10~PeV 
region, and are expected from the further operation of the experiment.
In addition, other experimental methods have to be 
developed to check the simulation codes even more rigorously. Such a 
stringent check consists of verifying absolute particle fluxes at 
ground level at energies where the primary flux is reasonably well known. 
Improvements in the interaction models are also under way. 
{\small NE}X{\small US} is 
in statu nascendi, a joint enterprise by the authors of VENUS and 
QGSJET \cite{werner}. It has become evident that a very precise 
description of the shower development in the atmosphere is needed if 
the mass of the primaries is to be estimated by means of ground 
level particle 
distributions.

\section{Acknowledgments}
The authors would like to thank the members of the engineering and
technical staff of the KASCADE 
collaboration who contributed with enthusiasm and engagement to the 
success of the experiment.

The Polish group gratefully acknowledges support by the Polish State 
Committee for Scientific Research (grant No. 2 P03B 16012).
The work has been partly supported by a grant of the Rumanian Ministry of
Research and Technology and by the research grant no. 94964 of the Armenian 
Government and ISTC project A 116. The support of the experiment by the 
Ministry for Research of the German Federal Government is gratefully 
acknowledged.

\end{document}